%% file: main.tex
\documentclass{article}
\usepackage{spconf,amsmath,amsfonts,graphicx}
\usepackage{mathtools}
\usepackage{pgfplots}
\usepackage{booktabs}
\usepackage{hyperref}
\usepackage{cleveref}
\usepackage{multirow}
\usepackage{amsthm}

\pgfplotsset{compat=newest}

\newtheorem{remark}{Remark}

\crefformat{section}{\S#2#1#3}
\crefformat{subsection}{\S#2#1#3}
\crefformat{subsubsection}{\S#2#1#3}

\title{Leveraging Heteroscedastic Uncertainty in Learning Complex Spectral Mapping for Single-channel Speech Enhancement}
\name{Kuan-Lin Chen$^{12\dagger}$, Daniel D. E. Wong$^{1*}$, Ke Tan$^1$, Buye Xu$^1$, Anurag Kumar$^1$, and Vamsi Krishna Ithapu$^1$\thanks{$\dagger$Work done during an internship at Meta Reality Labs Research.}\thanks{$^*$ The corresponding author.}}
\address{$^1$Meta Reality Labs Research\\$^2$Department of Electrical and Computer Engineering, University of California, San Diego}

\begin{document}
\ninept
\maketitle
\begin{abstract}
Most speech enhancement (SE) models learn a point estimate and do not make use of uncertainty estimation in the learning process. In this paper, we show that modeling heteroscedastic uncertainty by minimizing a multivariate Gaussian negative log-likelihood (NLL) improves SE performance at no extra cost. During training, our approach augments a model learning complex spectral mapping with a temporary submodel to predict the covariance of the enhancement error at each time-frequency bin. Due to unrestricted heteroscedastic uncertainty, the covariance introduces an undersampling effect, detrimental to SE performance. To mitigate undersampling, our approach inflates the uncertainty lower bound and weights each loss component with their uncertainty, effectively compensating severely undersampled components with more penalties. Our multivariate setting reveals common covariance assumptions such as scalar and diagonal matrices. By weakening these assumptions, we show that the NLL achieves superior performance compared to popular loss functions including the mean squared error (MSE), mean absolute error (MAE), and scale-invariant signal-to-distortion ratio (SI-SDR).
\end{abstract}
\begin{keywords}
Uncertainty, negative log-likelihood, neural networks, complex spectral mapping, speech enhancement
\end{keywords}

\input{introduction}
\input{proposed_methods}
\input{experimental_results}
\input{conclusion}

\bibliographystyle{IEEEbib_abbr}
\bibliography{refs}

\end{document}

%% file: introduction.tex
\section{Introduction} \label{sec:intro}
Speech enhancement (SE) aims at improving speech quality and intelligibility via recovering clean speech components from noisy recordings. It is an essential part of many applications such as teleconferencing \cite{hsu2022learning}, hearing aids \cite{pisha2019wearable}, and augmented hearing systems \cite{pisha2018wearable}. Modern SE approaches usually train a deep neural network (DNN) model to minimize a loss function on a target speech representation.
Because DNNs can be universal approximators \cite{cybenko1989approximation, hornik1989multilayer,telgarsky2016benefits} and capable of learning anything incentivized by the loss, designing DNN architectures on different target representations has been the most popular trend in SE.
Literature in this area is vast, \cite{lu2013speech,xu2013experimental,xu2014regression,wang2018supervised,tan2018convolutional,pandey2019tcnn,tan2019learning,hu20g_interspeech,hao2021fullsubnet,ijcai2022p582} for example.

Despite such progress, the most popular loss functions such as the MSE and MAE in SE make no or little use of uncertainty. We refer the reader to an excellent survey paper by Gawlikowski \textit{et al.} \cite{gawlikowski2021survey} for a thorough discussion of uncertainty in DNNs. From a probabilistic point of view, one can derive a loss function from a probabilistic distribution subject to certain constraints. For example, minimizing the MSE loss is equivalent to maximizing a Gaussian likelihood that assumes homoscedastic uncertainty, meaning that the variance associated with each squared error is a constant. The MAE loss also follows the same logic but with a Laplacian distribution. Taking complex spectral mapping \cite{fu2017complex,tan2019learning} for instance, optimizing the MSE or MAE on the complex spectrogram implicitly assumes a constant variance of the enhancement error on the real and imaginary parts at every time-frequency (T-F) bin. In fact, such an assumption even extends to speech signals in dissimilar noise conditions, e.g., different signal-to-noise ratios (SNRs). Although these loss functions are easy to use, they could limit the learning capability of a DNN model due to the underlying constant variance assumption.

The first effort in the literature training a neural network to minimize a Gaussian NLL dates back to a seminal work by Nix and Weigend \cite{nix1994estimating}. Although the Gaussian NLL has been earlier used in computer vision \cite{kendall2017uncertainties,amini2020deep}, its potential in SE remained unexplored until a recent work by Fang \textit{et al.} \cite{fang2022integrating}. They showed that a hybrid loss combining the SI-SDR \cite{le2019sdr,kolbaek2020loss} and the Gaussian NLL can outperform both MSE and SI-SDR losses. However, they also reported that minimizing a Gaussian NLL alone leads to inferior SE performance, highlighting the difficulty of using heteroscedastic uncertainty to improve perceptual scores in SE.

In this paper, we show that, at no extra cost in terms of compute, memory, and parameters, directly minimizing a Gaussian NLL yields significantly better SE performance than minimizing a conventional loss such as the MAE or MSE, and slightly better SE performance than the SI-SDR loss.
To the best of our knowledge, this is the first successful study that achieves improved perceptual metric performance by directly using heteroscedastic uncertainty for SE. Inspired by recent progress in uncertainty estimation \cite{seitzer2021pitfalls}, we reveal the main optimization difficulty and propose two methods: i) \textit{covariance regularization} and ii) \textit{uncertainty weighting} to overcome such a hurdle.
Experiments show that minimizing Gaussian NLLs using these methods consistently improves SE performance in terms of speech quality and objective intelligibility.

%% file: proposed_methods.tex
\section{Probabilistic Models and Assumptions} \label{sec:probabilistic_models_and_assumptions}
Let the received signal at a single microphone in the short-time Fourier transform (STFT) domain be $y_r^{t,f}+iy_i^{t,f}\in\mathbb{C}$ for all $(t,f)$ with the time frame index $t\in\{1,2,\cdots,T\}$ and frequency bin index $f\in\{1,2,\cdots,F\}$. Let $y\in\mathbb{R}^{2TF}$ be the vector representing every real part and imaginary part of the STFT representation of the received signal. We assume the clean signal is corrupted by additive noise, i.e., $y=x+v$ where $x$ and $v$ are the clean signal random vector and noise random vector, respectively, from the probabilistic perspective.
Now, we assume that the probability density of the clean signal given the received noisy signal and a conditional density model follows a multivariate Gaussian distribution
\begin{equation} \label{eq:posterior_distribution}
    p\left(x|y;\psi\right)=\frac{\exp\left (-\frac{1}{2}\left[x-\hat{\mu}_{\theta}(y)\right]^{\mathsf{T}}\hat{\Sigma}_{\phi}^{-1}(y)\left[x-\hat{\mu}_{\theta}(y)\right]\right)}{\sqrt{\left(2\pi\right)^n\det\hat{\Sigma}_{\phi}(y)}}
\end{equation}
where its conditional mean $\hat{\mu}_{\theta}(y)$ and covariance $\hat{\Sigma}_{\phi}(y)$ are directly learned from a dataset by a conditional density model $f_{\psi}$ parameterized by $\psi=\{\theta,\phi\}$ in supervised speech enhancement. Fig. \ref{fig:flowchart} illustrates the conditional density model $f_{\psi}$ and its difference compared to conventional SE models that only estimate clean speech. The map $f_{\psi}$ can be expressed as an augmented map consisting of an essential SE model $f_{\theta}$ and a temporary submodel $f_{\phi}$ such that
    \begin{equation}
         \begin{bmatrix}\hat{\mu}_{\theta}(y)\\\text{vec}\left[\hat{L}_{\phi}(y)\right]\end{bmatrix}=\begin{bmatrix}f_{\theta}(y)\\f_{\phi}(\tilde{y})\end{bmatrix}=f_{\psi}(y)
    \end{equation}
    where $f_{\psi}$, $f_{\theta}$, and $f_{\phi}$ are DNN models parameterized by $\psi$, $\theta$, and $\phi$, respectively. $\text{vec}[\cdot]$ is an operator that vectorizes a matrix. Because a valid covariance is symmetric positive semidefinite, the output of $f_{\phi}$ must be constrained to satisfy the property. To avoid imposing such a constraint, we design the map $f_{\phi}$ to estimate the lower Cholesky factor $\hat{L}_{\phi}$ of the covariance. The covariance can be later obtained by $\hat{\Sigma}_{\phi}(y)=\hat{L}_{\phi}(y)\hat{L}^{\mathsf{T}}_{\phi}(y)$. $\tilde{y}$ is the input feature of $f_{\phi}$ and a function of $y$, which can be $y$, an intermediate representation produced by $f_{\theta}$, or a combination of both. One can design different DNN architectures for $f_{\theta}$ and $f_{\phi}$. For example, $f_{\psi}$ can be an integrated DNN with two output branches, one for clean speech estimation and the other for covariance estimation, with shared weights between $f_{\theta}$ and $f_{\phi}$, i.e. $\theta\cap\phi\neq\emptyset$. Alternatively, $f_{\theta}$ and $f_{\phi}$ can be two separate DNNs, i.e. $\theta\cap\phi=\emptyset$. Below we point out key features of our framework.
    %The parameter set $\psi$ is optimized as a whole during training.
    \begin{remark} \label{remark:drop_submodel}
    The conditional mean $\hat{\mu}_{\theta}(y)$ is the enhanced signal so the submodel $f_{\phi}$ can be removed at inference time. Hence, one can use a much larger parameter set $\phi$ to design $f_{\phi}$ without increasing the complexity of the SE model at inference time.
    \end{remark}
    \begin{remark}
    The conditional covariance $\hat{\Sigma}_{\phi}(y)$ is also referred to as the uncertainty in this paper. Its homoscedasticity or heteroscedasticity is determined by assumptions made for the structure of the covariance. For example, the covariance can be assumed as a scalar matrix, a diagonal matrix, or a block diagonal matrix (see \cref{sec:nll}).
    \end{remark}
\begin{remark}
The form of conditional density in (\ref{eq:posterior_distribution}) can be obtained by assuming $x$ and $v$ are drawn from two multivariate Gaussian distributions. A \textit{Wiener filter} can be realized by estimating the mean and covariance of the joint distribution of $x$ and $v$. However, this approach requires a model to estimate more parameters, and such an extra cost cannot be removed at inference time.
\end{remark}

\input{flowchart}

Given a dataset $\{x_n,y_n\}_{n=1}^N$ containing pairs of target clean signal $x_n$ and received noisy signal $y_n$, we find the conditional mean $\hat{\mu}_{\theta}(y)$ and covariance $\hat{\Sigma}_{\phi}(y)$ maximizing the likelihood of the joint probability distribution
    $
        p(x_1,x_2,\cdots,x_N|y_1,y_2,\cdots,y_N;\psi)=\prod_{n=1}^Np\left(x_n|y_n;\psi\right)
    $
    where we assume the data points are independent and identically distributed (i.i.d.).
    
    \section{Multivariate Gaussian NLL} \label{sec:nll}
    Introducing the logarithmic function to the likelihood of the joint distribution and expanding terms according to (\ref{eq:posterior_distribution}), the maximization problem can be converted into minimizing the empirical risk using the following multivariate Gaussian NLL loss
    \begin{equation} \label{eq:multivariate_Gaussian_NLL}
        \ell^{\text{Full}}_{x,y}(\psi) = \left[x-\hat{\mu}_{\theta}(y)\right]^{\mathsf{T}}\hat{\Sigma}_{\phi}^{-1}(y)\left[x-\hat{\mu}_{\theta}(y)\right] + \log\det\hat{\Sigma}_{\phi}(y)
    \end{equation}
    of which the first term is an affinely transformed squared error between clean and enhanced speech, and the second term is a log-determinant term. Without imposing any assumptions on the covariance $\hat{\Sigma}_{\phi}$, the multivariate Gaussian NLL $\ell^{\text{Full}}_{x,y}(\psi)$ in (\ref{eq:multivariate_Gaussian_NLL}) uses a full matrix for the covariance. A full covariance matrix relaxes common assumptions such as uncorrelated real part and imaginary part at each T-F bin and uncorrelated T-F bins \cite{astudillo2009accounting,vincent2018audio}. Although $\ell^{\text{Full}}_{x,y}(\psi)$ is the most generalized formulation for a Gaussian NLL, the number of output units of the submodel $f_{\phi}$ is $4T^2F^2$, leading to exceedingly high training complexity. Assumptions (\cref{sec:homoscedastic_mse}, \cref{sec:heteroscedastic_uncorrelated_case}, and \cref{sec:block_diagonal}) are made to sparsify the covariance matrix, which in turn, reduces the complexity of the submodel and makes training amenable.
\subsection{Homoscedastic Uncertainty: An MSE Loss} \label{sec:homoscedastic_mse}
If the covariance $\hat{\Sigma}_{\phi}(y)$ is assumed to be a scalar matrix $\hat{\Sigma}_{\phi}(y)=cI$ where $c$ is a scalar constant and $I$ is an identity matrix, then we actually assume the uncertainty is homoscedastic. The log-determinant term in (\ref{eq:multivariate_Gaussian_NLL}) becomes a constant, and the affinely transformed squared error reduces to an MSE. In this case, minimizing the Gaussian NLL is equivalent to the empirical risk minimization using an MSE loss
$
    \ell^{\text{MSE}}_{x,y}(\theta)=\lVert x-\hat{\mu}_{\theta}(y)\rVert_2^2.
$
Apparently, the submodel $f_{\phi}$ is not needed for an MSE loss so the optimization is performed only on $\theta$. Many SE works fall into this category, e.g., \cite{lu2013speech,xu2013experimental,wang2018supervised,pandey2019tcnn,tan2019learning}.

\subsection{Heteroscedastic Uncertainty: A Diagonal Case} \label{sec:heteroscedastic_uncorrelated_case}
If every random variable in the random vector drawn from the conditional density $p(x|y)$ is assumed to be uncorrelated with the others, then the covariance reduces to a diagonal matrix. In this case, the Gaussian NLL ignores uncertainties across different T-F bins and between real and imaginary parts, leading to the following uncorrelated Gaussian NLL loss
\begin{equation} \label{eq:univariate_gaussian}
        \ell^{\text{Diagonal}}_{x,y}(\psi) = \sum_{t,f}\sum_{k\in\{r,i\}}\left[\frac{x_{k}^{t,f}-\hat{\mu}_{k;\theta}^{t,f}(y)}{\hat{\sigma}_{k;\phi}^{t,f}(y)}\right]^2+2\log \hat{\sigma}_{k;\phi}^{t,f}(y)
    \end{equation}
    where
    $\hat{\sigma}_{r;\phi}^{t,f}$ and $\hat{\sigma}_{i;\phi}^{t,f}$ denote the conditional standard deviation for the real and imaginary parts at $(t,f)$ bin, $\hat{\mu}_{r;\phi}^{t,f}$ and $\hat{\mu}_{i;\phi}^{t,f}$ denote their conditional means, and $x_{r}^{t,f}$ and $x_{i}^{t,f}$ denote the real and imaginary parts of $x$ at $(t,f)$ bin. In this case, the number of output units of the submodel $f_{\phi}$ is $2TF$. Note that the Gaussian NLL derived by Fang \textit{et al.} \cite{fang2022integrating} assumes circularly symmetric complex Gaussian distributions for both clean speech and noise. Such a circularly symmetric assumption is stronger than the assumption used in (\ref{eq:univariate_gaussian}). Consequently, their Gaussian NLL only has a variance term associated with each T-F bin whereas our formulation in (\ref{eq:univariate_gaussian}) allows the real and imaginary parts have their own variance.

\subsection{Heteroscedastic Uncertainty: A Block Diagonal Case} \label{sec:block_diagonal}
We relax the uncorrelated assumption imposed between every real and imaginary part in \cref{sec:heteroscedastic_uncorrelated_case} to take more uncertainty into account. In this case, the conditional covariance becomes a block diagonal matrix consisting of $2$-by-$2$ blocks, giving the Gaussian NLL loss
\begin{equation} \label{eq:bivariate_gaussian}
        \ell^{\text{Block}}_{x,y}(\psi) = \sum_{t,f}\underbrace{d_{\theta,x}^{t,f}(y)^{\mathsf{T}}
        \left[\hat{\Sigma}_{\phi}^{t,f}(y)\right]^{-1}
        d_{\theta,x}^{t,f}(y)+\log t_{\phi}^{t,f}(y)}_{z^{t,f}_{x,y}(\psi)}
    \end{equation}
    where
    $
        t_{\theta}^{t,f}(y)=\left[\hat{\sigma}_{r;\phi}^{t,f}(y)\hat{\sigma}_{i;\phi}^{t,f}(y)\right]^2-\left[\hat{\sigma}_{ri;\phi}^{t,f}(y)\right]^2,
    $
    $
        d_{\theta}^{t,f}(y)=\begin{bmatrix}
        x_{r}^{t,f}-\hat{\mu}_{r;\theta}^{t,f}(y)\\
        x_{i}^{t,f}-\hat{\mu}_{i;\theta}^{t,f}(y)
        \end{bmatrix},
    $
    $
        \hat{\Sigma}_{\phi}^{t,f}(y)=\begin{bmatrix}
        \left[\hat{\sigma}_{r;\phi}^{t,f}(y)\right]^2& \hat{\sigma}_{ri;\phi}^{t,f}(y)\\
        \hat{\sigma}_{ri;\phi}^{t,f}(y) & \left[\hat{\sigma}_{i;\phi}^{t,f}(y)\right]^2
        \end{bmatrix},
    $
    and $\hat{\sigma}_{ri;\phi}^{t,f}$ is the covariance between the real and imaginary parts at $(t,f)$ bin.
    Compared to the uncorrelated case in \cref{sec:heteroscedastic_uncorrelated_case}, the submodel $f_{\phi}$ needs to additionally predict one more parameter at every $(t,f)$ bin, resulting in a submodel with $3TF$ output units, while the inference-time complexity of the SE model $f_{\theta}$ remains the same as using an MSE loss or uncorrelated Gaussian NLL loss (Remark \ref{remark:drop_submodel}).
\section{On Mitigating Undersampling} \label{sec:undersampling}
The optimization difficulty in minimizing a Gaussian NLL can be revealed by its average first-order derivative. Taking the uncorrelated Gaussian NLL for example, the expected first-order derivative of $\ell^{\text{Digonal}}_{x,y}$ with respect to $\hat{\mu}_{r;\theta}^{t,f}$ can be approximated by
    \begin{equation} \label{eq:gradient_mean}
        \mathbb{E}_{x,y}\left[\frac{\partial\ell^{\text{Diagonal}}_{x,y}}{\partial\hat{\mu}_{r;\theta}^{t,f}}\right]\approx\frac{2}{N}\sum_{n=1}^N\frac{\hat{\mu}_{r;\theta}^{t,f}(y_n)-x_{n,r}^{t,f}}{\left[\hat{\sigma}_{r;\phi}^{t,f}(y_n)\right]^2}.
    \end{equation}
    Given the unconstrained variance in the denominator, a larger variance makes the model $f_{\theta}$ harder to converge to a clean component compared to a loss component with a smaller variance. This undersampling issue was pointed out in a recent work by Seitzer \textit{et al.} \cite{seitzer2021pitfalls}, in which they proposed the $\beta$-NLL to mitigate undersampling. However, the $\beta$-NLL was only developed for the univariate Gaussian NLL. It can be used in (\ref{eq:univariate_gaussian}) because the uncorrelated multivariate Gaussian NLL can be decomposed into many univariate Gaussian NNLs, while (\ref{eq:bivariate_gaussian}) can only be decomposed into many bivariate Gaussian NLLs, which requires a more generalized approach.% The following two techniques are proposed to mitigate undersampling in minimizing a general multivariate Gaussian NLL.
\subsection{Covariance Regularization}
Let $\delta>0$ be the lower bound of the eigenvalues of the Cholesy factor of the covariance matrix. The output of $f_{\phi}$ is modified by
\begin{equation}
    \left[\hat{L}_{\phi}^{\delta}(y)\right]_{mm}=\max\left\{\left[\hat{L}_{\phi}(y)\right]_{mm},\delta\right\}.
\end{equation}
for all $m\in\{1,2,\cdots,2TF\}$ where $\hat{L}_{\phi}^{\delta}(y)$ is now the regularized output of $f_{\phi}$.
As the degree of undersampling is affected by the ratio of the largest variance to the smallest variance, suitably increasing $\delta$ can reduce the ratio and hence mitigate undersampling. However, a large $\delta$ can saturate uncertainties, driving the NLL toward the MSE.
\subsection{Uncertainty Weighting}
Because a large variance can make the gradient of a loss component small, assigning a larger weight for such a loss component would alleviate undersampling. This is the intuition of $\beta$-NLL. To extend it to a multivariate Gaussian NLL, we propose an \textit{uncertainty weighting} approach, which assigns a larger weight for a loss component according to the \textit{minimum eigenvalue} of the covariance matrix. Applying such a strategy to (\ref{eq:bivariate_gaussian}) leads to the following loss function
\begin{equation}
\ell^{\beta\text{-Block}}_{x,y}(\psi) =\sum_{t,f}\lambda_{\text{min}}\left[\hat{\Sigma}_{\phi}^{t,f}(y)\right]^{\beta}z^{t,f}_{x,y}(\psi)
\end{equation}
where $\lambda_{\text{min}}\left[\cdot\right]$ gives the minimum eigenvalue which is treated as a constant. No gradients are propagated through $\lambda_{\text{min}}\left[\cdot\right]$. $\beta\in[0,1]$ is a hyperparameter controlling the degree of uncertainty weighting. When $\beta=0$, $\ell^{\beta\text{-Block}}_{x,y}(\psi)$ reduces to $\ell^{\text{Block}}_{x,y}(\psi)$.%When $\beta=1$, the gradient of $\ell^{\beta\text{-Block}}_{x,y}(\psi)$ with respect to the mean can get more aligned to the gradient of the MSE.
To mitigate undersampling while exploiting heteroscedastic uncertainty, we pick $\beta=0.5$, which is is also the suggested value of $\beta$-NLL.

%% file: flowchart.tex
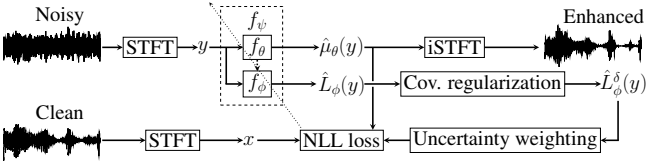
\begin{figure}[tb]
    \centering
    \resizebox{\columnwidth}{!}{\begin{tikzpicture}
\begin{axis}[
    xmin=0, xmax=1,
    ymin=-1, ymax=1,
    hide axis
    ]
    \addplot[smooth] table{noisy_example.txt};
\end{axis}
\begin{scope}[xshift=0cm, yshift=-3.35cm]
\begin{axis}[
    xmin=0, xmax=1,
    ymin=-1.8, ymax=1.8,
    hide axis
    ]
    \addplot[smooth] table{clean_example.txt};
\end{axis}
\end{scope}
\node (rect) at (5.2,2.8) [draw,thick,minimum width=1.0cm,minimum height=0.5cm] {\huge STFT};
\draw [-stealth,ultra thick](3.5,2.8) -- (4.23,2.8);
\draw [-stealth,ultra thick](6.16,2.8) -- (6.9,2.8);
\node (rect) at (9,2.8) [draw,thick,minimum width=1cm,minimum height=0.5cm] {\huge $\huge f_{\theta}$};
\draw [-stealth,ultra thick](7.3,2.8) -- (8.5,2.8);
\node (rect) at (9.0,1.5) [draw,thick,minimum width=1cm,minimum height=0.5cm] {\huge $f_{\phi}$};
\draw [-stealth,dotted, ultra thick](9.00,2.35) -- (9.00,1.95);
\draw [-,ultra thick](7.9,2.8) -- (7.9,1.5);
\draw [-stealth,ultra thick](7.872,1.5) -- (8.5,1.5);

\draw [-,ultra thick] (13.1,2.8) to (13.1,1.6);
\draw [-stealth,ultra thick] (13.1,1.4) to (13.1,-0.13);

\node at (12,2.8) [draw=none,thick,] {\huge $\hat{\mu}_{\theta}(y)$};
\draw [-stealth,ultra thick](9.5,2.8) -- (11.2,2.8);
\node at (12.0,1.5) [draw=none,thick] {\huge $\hat{L}_{\phi}(y)$};
\draw [-stealth,ultra thick](9.5,1.5) -- (11.1,1.5);
\node (rect) at (17.0,1.5) [draw,thick,minimum width=1cm,minimum height=0.5cm] {\huge Cov. regularization};
\draw [-stealth,ultra thick](12.8,1.5) -- (14.07,1.5);
\node at (22.0,1.5) [draw=none,thick] {\huge $\hat{L}_{\phi}^{\delta}(y)$};
\draw [-stealth,ultra thick](19.90,1.5) -- (21.2,1.5);

\node at (2,3.9) [draw=none,thick] {\huge Noisy};
\node at (2,0.5) [draw=none,thick] {\huge Clean};
\node at (21.3,3.9) [draw=none,thick] {\huge Enhanced};
\node (rect) at (6,-0.5) [draw,thick,minimum width=1.0cm,minimum height=0.5cm] {\huge STFT};
\draw [-stealth,ultra thick](3.5,-0.5) -- (5.04,-0.5);
\node at (8.7,-0.5) [draw=none,thick] {\huge $x$};
\node at (7.08,2.8) [draw=none,thick] {\huge $y$};
\draw [-stealth,ultra thick](6.96,-0.5) -- (8.5,-0.5);

\node (rect) at (8.8,2.5) [draw, dashed, thick,minimum width=2.2cm,minimum height=3.5cm] {};
\node at (9.0,3.7) [draw=none,thick] {\huge $f_{\psi}$};

\node (rect) at (17.8, -0.5) [draw,thick,minimum width=1cm,minimum height=0.5cm] {\huge Uncertainty weighting};
\draw [-,ultra thick](21.8,0.9) -- (21.8,-0.5);
\draw [-stealth,ultra thick](21.828,-0.5) -- (21.18,-0.5);

\node (rect) at (12.0,-0.5) [draw, thick,minimum width=2.5cm,minimum height=0.5cm] {\huge NLL loss};
\draw [-stealth,ultra thick](9.0,-0.5) -- (10.52,-0.5);
\draw [-stealth,ultra thick](14.42,-0.5) -- (13.48,-0.5);
\draw [-stealth,dotted,thick](10.55,-0.05) -- (7.3,4.4);

\node (rect) at (16,2.8) [draw,thick,minimum width=1.0cm,minimum height=0.5cm] {\huge iSTFT};
\draw [-stealth,ultra thick](12.8,2.8) -- (14.93,2.8);

\begin{scope}[xshift=19.2cm, yshift=0cm]
\begin{axis}[
    xmin=0, xmax=1,
    ymin=-1, ymax=1,
    hide axis
    ]
    \addplot[smooth] table{enhanced_example.txt};
\end{axis}
\end{scope}
\draw [-stealth,ultra thick](17.06,2.8) -- (19.0,2.8);
\end{tikzpicture}}
    \caption{We augment an SE model $f_{\theta}$ with a temporary submodel $f_{\phi}$ to estimate heteroscedastic uncertainty during training. The augmented model $f_{\psi}$ is trained to minimize a multivariate Gaussian NLL (see \cref{sec:nll}). The methods of covariance regularization and uncertainty weighting are proposed to overcome the optimization difficulty that arises in optimizing the NLL (see \cref{sec:undersampling}).}
    \label{fig:flowchart}
\end{figure}

%% file: experimental_results.tex
\section{Experiments} \label{sec:experimental_results}
\begin{table*}[t]
            \centering
            \begin{tabular}{cllrrr|rrr|rrr|rrr}
                \toprule
                & $\delta$& $\beta$ & \multicolumn{3}{c}{WB-PESQ} &  \multicolumn{3}{c}{STOI (\%)} & \multicolumn{3}{c}{SI-SDR (dB)} & \multicolumn{3}{c}{NORESQA-MOS}\\
                SNR (dB)&&& -5 & 0 & 5 &-5 & 0 & 5 &-5 & 0 & 5 & -5 & 0 & 5 \\
                \midrule
                Unprocessed&\multicolumn{2}{c}{n/a}& 1.11 & 1.15 & 1.24 & 69.5 & 77.8 & 85.2 & -5.00 & 0.01 & 5.01 & 2.32 & 2.36 & 2.45 \\
                \midrule
                MAE & & & 1.50 & 1.76 & 2.09 & 84.4 & 90.4 & 93.9 & 9.83  & 12.63 & 15.02& 2.77 & 3.27 & 3.65 \\
 		        MSE & \multicolumn{2}{c}{n/a} & 1.63 & 1.94 & 2.29 & 85.1 & 90.6 & 94.0 & 10.24  & 13.21 & 15.97& 2.86 & 3.52 & 4.02 \\
 		        SI-SDR & &  & 1.71 & 2.04 & 2.42 & 86.5 & 91.5 & 94.6 &\textbf{10.96} & \textbf{13.92} & \textbf{16.80} & 3.05 & 3.65 & 4.20 \\
                \midrule
                \multirow{ 3}{*}{\shortstack{Gaussian NLL: \\ Diagonal $\hat{\Sigma}_{\phi}$}}& $0.0001$&$0$ & 1.11 & 1.18 & 1.28 & 69.6 & 77.3 & 83.0 & 0.79 & 4.37 & 7.48 & 1.95	& 2.16 & 2.40\\
                & $0.01$&$0$ & 1.59 & 1.88 & 2.28 & 83.5 & 89.7 & 93.7 & 7.65 & 10.61 & 13.31 & 2.97 & 3.60 & 4.14\\
                & $0.01$&$0.5$ & 1.74 & 2.08 & 2.48 & 86.2 & 91.3 & 94.6 & 9.83 & 12.55 & 15.01 & 3.14 & 3.77 & 4.25\\
                \midrule
                \multirow{ 7}{*}{\shortstack{Gaussian NLL: \\Block diagonal $\hat{\Sigma}_{\phi}$}} &$0.0001$&$0$ & 1.07 & 1.08 & 1.11 & 59.4 & 66.5 & 72.0 & -6.46 & -4.20 & -2.82 & 1.56 & 1.47 & 1.44 \\
                &$0.001$&$0$ & 1.53 & 1.80 & 2.19 & 82.6 & 89.1 & 93.3 & 7.08 & 10.08 & 13.01 & 2.71 & 3.33 & 3.97 \\
                &$0.01$&$0$ & 1.61 & 1.92 & 2.33 & 83.9 & 90.1 & 94.0 & 7.82 & 10.73 & 13.51 & 2.98 & 3.60 & 4.15 \\
 		        & $0.001$&$0.5$ & 1.73 & 2.08 & 2.49 & 86.0 & 91.4 & 94.7 & 9.71 & 12.62 & 15.41 & 3.11 & 3.79 & 4.30\\
 		        & $0.005$&$0.5$ & \textbf{1.75} & \textbf{2.11} & \textbf{2.52} & 86.4 & 91.6 & 94.8 & 10.09 & 13.05 & 15.88 & 3.07 & 3.75 & 4.22\\
 		        &$0.01$&$0.5$ & \textbf{1.75} & 2.10 & 2.50 & \textbf{86.7} & \textbf{91.8} & \textbf{94.9} & 10.22 & 13.15 & 15.99 & \textbf{3.23} & \textbf{3.89} & \textbf{4.35} \\
 		        & $0.05$&$0.5$ & 1.72 & 2.08 & 2.49 & 86.3 & 91.6 & 94.8 & 10.12 & 13.09 & 15.86 & 2.96 & 3.63 & 4.15\\
                \bottomrule
            \end{tabular}
            \caption{The methods of covariance regularization and uncertainty weighting effectively improve perceptual metric performance of multivariate Gaussian NLLs. The NLL using a block diagonal covariance with suitable $\delta$ and $\beta$ outperforms the MAE, MSE, and SI-SDR.} \label{tab:correlated_cov_assumption_uncertainty_weighting}
\end{table*}
The DNS dataset \cite{reddy21_interspeech} is used as the corpus for all experiments. By randomly mixing the speech and noise signals in the DNS dataset, we simulate our training, validation, and test sets, which consist of 500K, 1K, and 1.5K pairs of noisy and clean utterances, respectively. The SNR for each noisy utterance in the training and validation sets is randomly sampled between -5 and 5 dB. For the test set, -5, 0, and 5 dB SNRs are used, equally dividing the 1.5K utterances. Note that all test speakers are excluded from the training and validation sets and all utterances are sampled at 16 kHz, each of which is truncated to 10 seconds. The window size and hop size of STFT are 320 and 160 points, respectively, where the Hann window is used.
We adopt the gated convolutional recurrent network (GCRN) \cite{tan2019learning} as the SE model $f_{\theta}$ for investigation. Given that the original GCRN has an encoder-decoder architecture with long short-term memory (LSTM) in between, we formulate the temporary submodel $f_{\phi}$ as an additional decoder that takes the output of the in-between LSTM as input. Hence the augmented model $f_{\psi}$ formed by these two models is a GCRN with two distinct decoders. For comparison, we train three SE models $f_{\theta}$ individually using the MAE, MSE, and SI-SDR loss ($\ell^{\text{SI-SDR}}$), respectively. Another model $f_{\psi}$ is trained with the Gaussian NLL. At inference time, $f_{\psi}$ drops $f_{\phi}$, so all SE models for comparison have the same DNN architecture and number of parameters. The Adam \cite{kingma2015adam} optimizer is adopted to train all the models. The learning rate is $0.0004$ and the batch size is 128. All models are trained for 300 epochs. After each training epoch, the model is evaluated on the validation set, and the best model is determined by the validation results. We measure SE performance using multiple metrics, including wideband perceptual evaluation of speech quality (WB-PESQ) \cite{rix2001perceptual}, short-time objective intelligibility (STOI) \cite{taal2011algorithm} (\%), SI-SDR \cite{le2019sdr} (dB), and NORESQA-MOS \cite{noresqamos} on the test set.

\noindent\textbf{Calibration of the Probabilistic Model:}
\begin{figure}[tb]
    \begin{minipage}[b]{0.5\linewidth}
  \centering
  \centerline{\includegraphics[width=4.0cm]{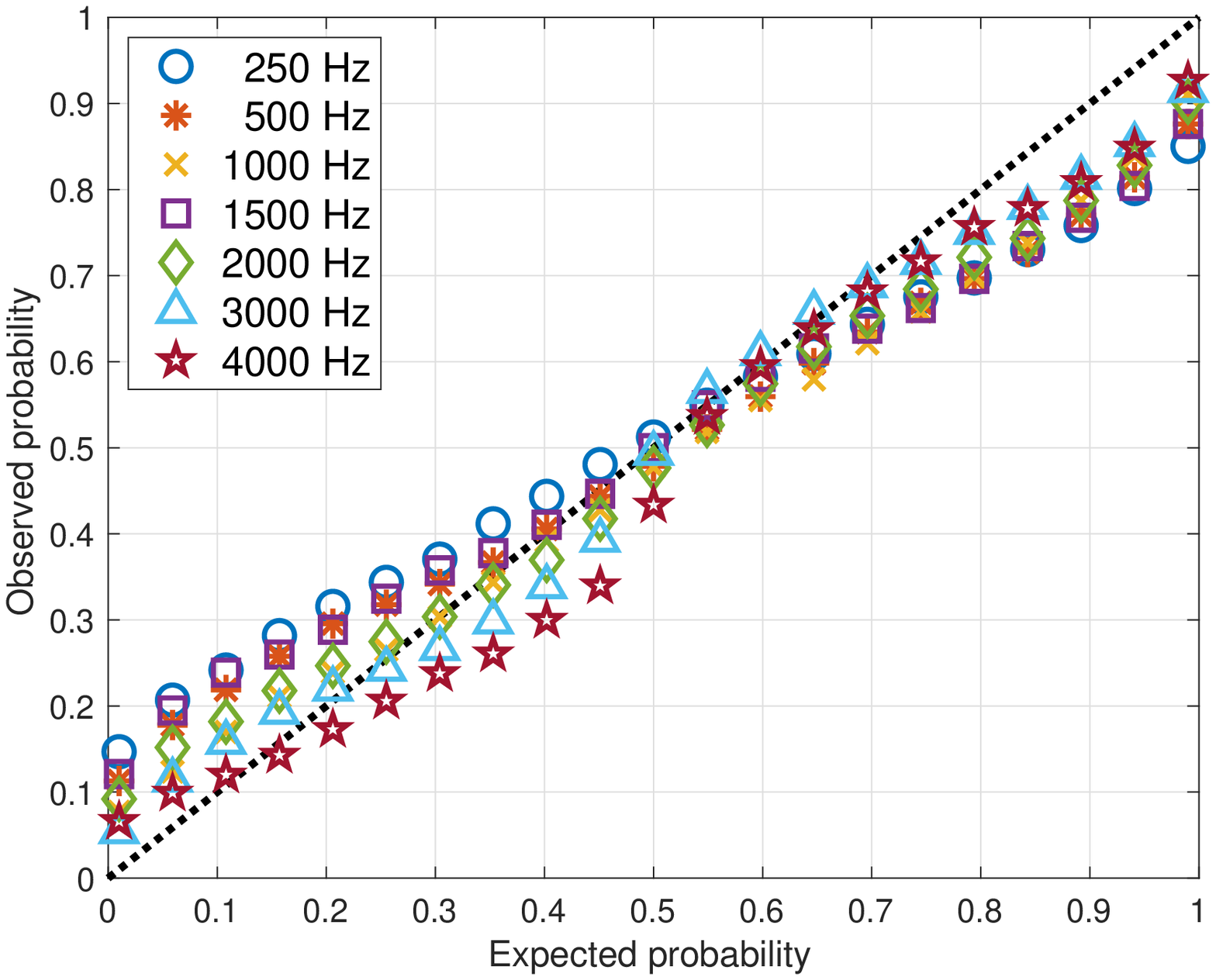}}
  \centerline{(a) Real part.}
\end{minipage}
\begin{minipage}[b]{.5\linewidth}
  \centering
  \centerline{\includegraphics[width=4.0cm]{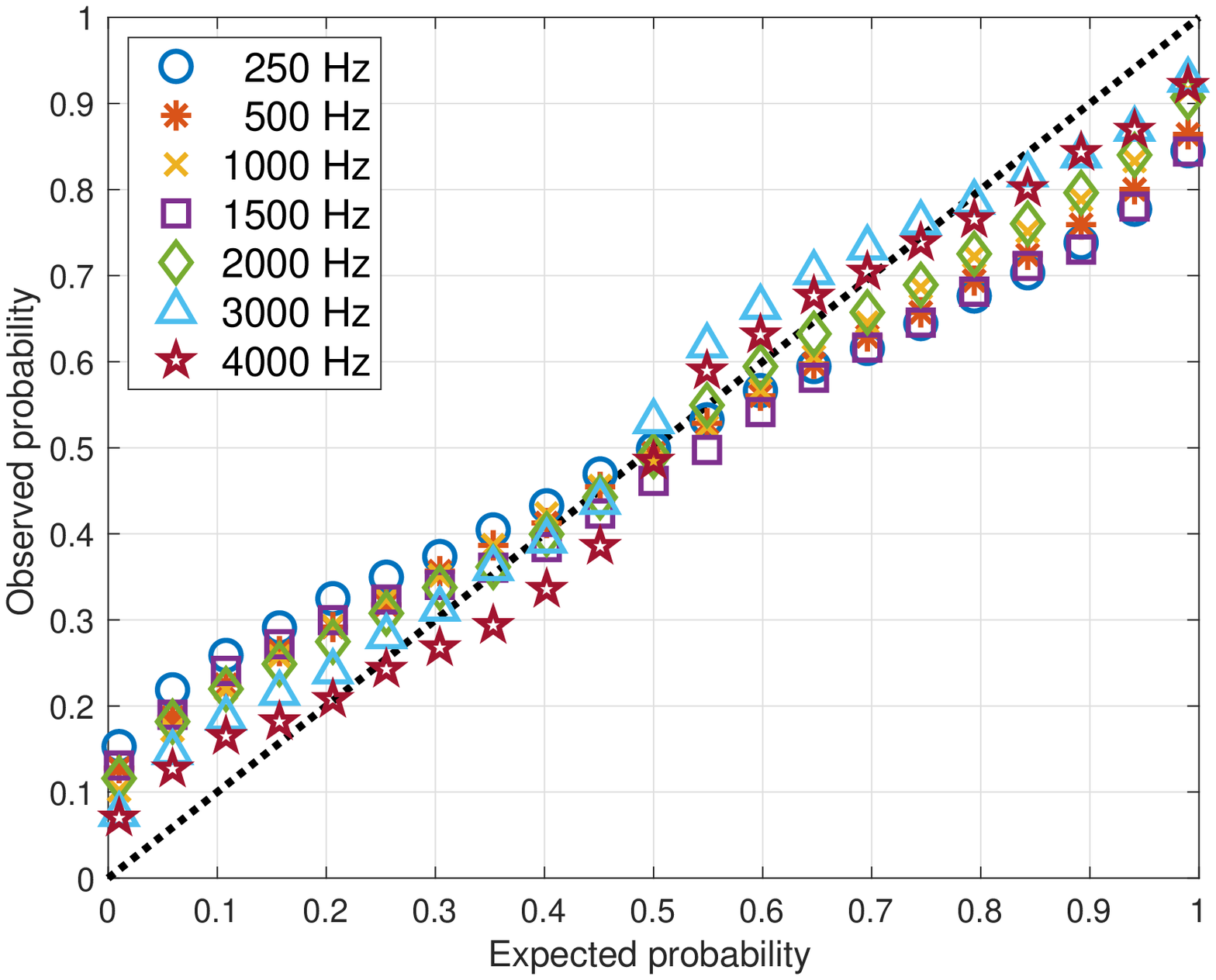}}
  \centerline{(b) Imaginary part.}
\end{minipage}
    \caption{The Q-Q plots suggest that the predictive Gaussian distributions reasonably capture the populations of the clean speech.}
    \label{fig:quantile_quantile_plot_delta1e-2_beta05}
\end{figure}
Each quantile-quantile (Q-Q) plot at a different frequency component in Fig. \ref{fig:quantile_quantile_plot_delta1e-2_beta05}(a) compares the population of a clean speech with the predictive Gaussian distribution for the real part on a frequency component. The Q-Q plots for the imaginary part are shown in Fig. \ref{fig:quantile_quantile_plot_delta1e-2_beta05}(b). All these Q-Q plots are close to the main diagonal, showing that calibration qualities seem to be acceptable. The predictive distribution is obtained by training an SE model with the Gaussian NLL $\ell^{\beta\text{-Block}}$ using $\delta=0.01$ and $\beta=0.5$ and feeding a random noisy utterance at $5$ dB SNR from the test set to the model. One can probably argue that such calibration qualities are sufficient to achieve SE performance improvements.

\noindent\textbf{Covariance Regularization and Uncertainty Weighting:}
Table \ref{tab:correlated_cov_assumption_uncertainty_weighting} shows that the eigenvalue lower bound of the lower Cholesky factor $\delta$ plays an important role in minimizing Gaussian NLLs. When $\delta=0.0001$, we find that it is very difficult to optimize the Gaussian NLL, and the trained model completely fails to enhance speech. Such an issue can be significantly improved by increasing $\delta$. Taking the block diagonal case for example, $\delta=0.001$ gives an SE model with reasonable perceptual metric performance, and further increasing $\delta$ to $0.01$ gives even better performance.
On the other hand, Table \ref{tab:correlated_cov_assumption_uncertainty_weighting} shows that applying the uncertainty weighting method with $\beta=0.5$ consistently improves SE performance for different eigenvalue lower bounds $\delta$ and covariance assumptions.

\begin{table}[t]
\centering
\setlength{\tabcolsep}{2.6pt} % let TeX compute the intercolumn space
\begin{tabular}{cccc|ccc|ccc}
\toprule
&\multicolumn{3}{c}{WB-PESQ} & \multicolumn{3}{c}{STOI (\%)} & \multicolumn{3}{c}{SI-SDR (dB)} \\
SNR (dB) &-5 & 0 & 5 & -5 & 0 & 5 & -5 & 0 & 5 \\
\midrule
%Unprocessed & \multicolumn{2}{c}{n/a} & \multicolumn{2}{c}{n/a} & 1.11 & 1.15 & 1.24 & 69.5 & 77.8 & 85.2 & -5.00 & 0.01 & 5.01 \\
Hybrid& 1.77 & 2.14 & 2.53 & 86.9 & 91.9 & 94.9 & 10.62 & 13.58 & 16.30 \\
\bottomrule
\end{tabular}
\caption{Performance evaluation of a hybrid loss defined by $\ell^{\text{Hybrid}}=\alpha\ell^{\beta\text{-Block}}+(1-\alpha)\ell^{\text{SI-SDR}}$ with $\alpha=0.99$, $\delta=0.01$, and $\beta=0.5$.}
\label{tab:hybrid}
\end{table}

\noindent\textbf{Covariance Assumptions:}
Table \ref{tab:correlated_cov_assumption_uncertainty_weighting}  also shows that the Gaussian NLL using the block diagonal covariance assumption outperforms the Gaussian NLL using the diagonal covariance assumption for both $\beta=0$ and $\beta=0.5$ under $\delta=0.01$. These improvements show that modeling more heteroscedastic uncertainty is beneficial for SE. Note that modeling more uncertainty implicitly relaxes more assumptions for the loss function, which gives the SE model more flexibility to learn better complex spectral mapping.

\noindent\textbf{In Comparison to Losses without Exploiting Uncertainty:}
Table \ref{tab:correlated_cov_assumption_uncertainty_weighting} shows that the Gaussian NLL using the block diagonal covariance with $\delta=0.01$ and $\beta=0.5$ substantially outperforms the MAE and MSE loss functions that assume homoscedastic uncertainty. The Gaussian NLL also slightly outperforms the SI-SDR loss.
To determine if the Gaussian NLL gives statistically significant improvements over the SI-SDR loss, we perform the paired Student's $t$-test that assumes the two-tailed distribution.
The $p$-values for WB-PESQ, STOI, and NORESQA-MOS are much less than $0.1\%$, implying that these improvements are statistically significant.
It should be noted that, although the SI-SDR loss yields competitive perceptual metric performance, it does not preserve the level of the clean speech signal due to its scale invariance, which would require additional rescaling in real applications. In contrast, the proposed Gaussian NLL preserves the level of clean speech while achieving superior perceptual metric performance.

\noindent\textbf{A Hybrid Loss:}
Table \ref{tab:hybrid} shows that a hybrid loss gives better performance than every single-task loss in Table \ref{tab:correlated_cov_assumption_uncertainty_weighting}, suggesting that combining the Gaussian NLL with the SI-SDR loss is indeed beneficial. Such a result supports a multi-task learning strategy for SE.

%% file: conclusion.tex
\section{Conclusion}
In this study, we have developed a novel framework to improve SE performance by modeling uncertainty in the estimation. Specifically, we jointly optimize an SE model to learn complex spectral mapping and a temporary submodel to minimize a multivariate Gaussian NLL.
In our multivariate setting, we reveal common covariance assumptions and propose to use a block diagonal assumption to leverage more heteroscedastic uncertainty for SE. To overcome the optimization difficulty induced by the multivariate Gaussian NLL, we propose two methods, covariance regularization and uncertainty weighting, to mitigate the undersampling effect. With these methods, the multivariate Gaussian NLL substantially outperforms conventional losses including the MAE and MSE, and slightly outperforms the SI-SDR. To our best knowledge, this study is the first to show that directly minimizing a Gaussian NLL can improve SE performance, with our approach. Furthermore, such improvements in SE are achieved without extra computational cost at inference time.